\documentclass[12pt]{article}
\usepackage{amsfonts,amssymb}
\usepackage{mathrsfs}

\newcommand{\R}{{\mathbb{R}}}

\newcommand{\C}{{\mathbb{C}}}

\newcommand{\beq}{\begin{equation}}
\newcommand{\eeq}{\end{equation}}
\newcommand{\bea}{\begin{eqnarray}}
\newcommand{\eea}{\end{eqnarray}}
\newcommand{\ra}{\rightarrow}

\newcommand{\cd}{\partial}

\newcommand{\wh}{\widehat}

\newcommand{\ip}[1]{\langle#1\rangle}

\def \dstar{\delta}
\def \d{\mathrm{d}}

\newcommand{\vol}{{\rm vol}}

\newcommand{\eps}{\varepsilon}

\newcommand{\area}{\mathsf{Area}}

\renewcommand{\phi}{\varphi}

\begin{document}

\title{The ground state energy of a charged particle on a Riemann surface}
\author{
J.M. Speight\thanks{E-mail: {\tt speight@maths.leeds.ac.uk}}\\
School of Mathematics, University of Leeds\\
Leeds LS2 9JT, England
}

\date{}
\maketitle

\begin{abstract}
It is shown that the quantum ground state energy of particle of
mass $m$ and electric charge $e$ moving on a compact Riemann surface under
the influence of a constant magnetic field of strength $B$ is
$E_0=\frac{eB}{2m}$. Remarkably, this formula is completely independent of
both the geometry and topology of the Riemann surface. The formula is
obtained by reinterpreting the quantum Hamiltonian as the second variation
operator of an associated classical variational problem.
\vspace*{0.3cm}\newline
\end{abstract}

\maketitle

Consider a point particle of mass $m$ and electric charge $e$ moving on a
compact Riemann surface $\Sigma$ under the influence of a uniform
magnetic field of strength $B$. The purpose of this letter is to show
that the ground state energy of such a particle, in 
nonrelativistic quantum mechanics, is 
\beq\label{e0}
E_0=\frac{eB}{2m}
\eeq
where we have chosen to use natural units ($\hbar=c=1$). The remarkable
thing about this formula is that it is completely independent of the choice
of surface $\Sigma$; not only is it independent of the {\em metric} on $\Sigma$,
and hence of local details of the shape of $\Sigma$,
it is also independent of the {\em genus} of $\Sigma$. 

The equivalent problem on euclidean $\R^2$ is, of course, well understood
\cite{lanlif}, the whole energy spectrum being easily computed,
\beq
E_n^{(\R^2)}=\left(n+\frac12\right)\frac{eB}{m}, \quad n=0,1,2,\ldots.
\eeq
Note that $E_0^{(\R^2)}$, known in condensed matter contexts as
the energy of the {\em first Landau level},  coincides precisely with the 
ground state energy on a compact domain, (\ref{e0}). So compactifying space
leaves the ground state energy completely unchanged. This seems to be
a special property of just $E_0$ which does not hold for $E_n$ with $n\geq 1$.
Indeed, in the case where $\Sigma=S^2$ with the round metric of radius $R$,
Haldane \cite{hal}
has exploited the $SO(3)$ symmetry to obtain the full spectrum
\beq
E_n^{(S^2)}=\left(n+\frac12\right)\frac{eB}{m}+\frac{n(n+1)}{2mR^2}, \quad n=0,1,2,\ldots,
\eeq
which, for $n\geq 1$, agrees with $E_n^{(\R^2)}$ only in the limit $R\ra\infty$.
So there is something special about $E_0$ which protects it against change
even under topology-changing deformations of the domain. It would be 
interesting to see whether the same formula for $E_0$ holds on
arbitrary complete oriented two-manifolds, without the assumption of 
compactness. The example of $\Sigma=\R^2$ suggests it may, although our argument
relies strongly on compactness of $\Sigma$.

Our method 
is to reinterpret the quantum Hamiltonian as (part of) the second variation
of the energy of a related gauge theory, then use a known phase transition
in this field theory to deduce the lowest eigenvalue. 
The phase transition is analogous to that which occurs in a type II
superconductor at the upper critical magnetic field $H_{c2}$, where
the normal state becomes stable and energetically
preferred over the Abrikosov vortex lattice \cite{tin}.
To the best of our
knowledge this is the first time that the logic of the stability analysis
has been inverted in this fashion: usually one uses spectral properties
of a differential operator, possibly reinterpreted as a quantum Hamiltonian,
to deduce stability properties of the classical system, whereas we argue
in exactly the opposite direction.

We begin by defining the quantum Hamiltonian in local coordinates.
Let $x_1,x_2$ be isothermal local coordinates on $\Sigma$, so that the
metric is locally
\beq
g=\Omega(x_1,x_2)^2(dx_1^2+dx_2^2)
\eeq
for some smooth function $\Omega$. The magnetic field is
$B=\Omega^{-2}(\cd_1A_2-\cd_2A_1)$ where $A=A_1\d x_1+A_2\d x_2$ is a 
local gauge potential.
The quantum Hamiltonian of a particle of mass $m$ and electric charge $e$
moving on $\Sigma$ in this background field is
\beq
H\psi=-\frac{1}{2m\Omega^2}(\cd_i-ieA_i)(\cd_i-ieA_i)\psi.
\eeq
It is this operator, in the case where $B$ is constant, whose lowest eigenvalue
we claim is $E_0$, as in (\ref{e0}).

To proceed further, it is convenient to formulate things in a global, coordinate
free language. If $B\neq 0$ then, since $\Sigma$ is compact, one should
not think of $A$ as a one-form on $\Sigma$, but rather as the local
coordinate expression of a metric connexion $\nabla$ on a hermitian 
line bundle
$(L,h)$ over $\Sigma$. The wave function $\psi$ is not a
mapping $\Sigma\ra \C$, but rather a section of $L$. 
Explicitly, let $h$ be the fibre metric on $L$ and
$\eps$ be a local unit section of $L$ (that is, $|\eps|^2=h(\eps,\eps)=1$).
Then the connexion $\nabla$ acts on an arbitrary local section $\phi=f\eps$ as
\beq
\nabla_X(f\eps)=(X[f]-ieA(X)f)\eps,
\eeq
where $X\in T_p\Sigma$.
Reality of $A$ ensures that $\nabla$ is metric compatible, that is,
$X[h(\phi,\psi)]=h(\nabla_X\psi,\psi)+h(\psi,\nabla_X\phi)$. Associated
with $\nabla$ are an exterior differential operator $\d^\nabla:\Omega^p(L)\ra
\Omega^{p+1}(L)$ and its $L^2$ adjoint (the coderivative) $
\dstar^\nabla:\Omega^p(L)\ra
\Omega^{p-1}(L)$, where $\Omega^p(L)$ denotes the space of $p$-forms on
$\Sigma$ taking values in $L$. Explicitly, given any
$\phi\in\Gamma(L)$ and $\lambda\in\Omega^p(\Sigma)$,
\beq
(\d^\nabla\phi)(X)=\nabla_X\phi,\qquad
\d^\nabla(\phi\lambda)=(\d^\nabla\phi)\wedge\lambda+\phi\d\lambda,
\eeq
and $\dstar^\nabla=-*\d^\nabla *$ where $*$ is the Hodge isomorphism
$\Omega^p(\Sigma)\ra\Omega^{2-p}(\Sigma)$ induced by the metric $g$. One sees immediately
that in this language
\beq
H\psi=\frac{1}{2m}\dstar^\nabla\d^\nabla\psi=\frac{1}{2m}\Delta^\nabla\psi
\eeq
where $\Delta^\nabla$ denotes the natural laplacian operator
on $(L,h,\nabla)$. 
This laplacian
is manifestly
non-negative, and is known to be elliptic \cite{atibot}, so
its spectrum is discrete, non-negative, and each eigenvalue has
finite multiplicity. Hence, it has a lowest eigenvalue $\lambda_0\geq 0$,
and $E_0=\frac{\lambda_0}{2m}$. 

It is not hard to show that $\lambda_0\geq eB$.
The curvature $F^\nabla$ of $\nabla$ is
$\d^\nabla\d^\nabla\in\Omega^2(End(L))$ which can be identified globally
with an imaginary 2-form, coinciding locally with
\beq
F^\nabla=-ie\d A=-ieB\vol_\Sigma
\eeq
where $\vol_\Sigma$ is the volume form on $(\Sigma,g)$. It is well known
that
\beq
n=\int_\Sigma\frac{iF^\nabla}{2\pi}=\frac{e}{2\pi}\int_\Sigma B\vol_\Sigma
\eeq
is an integer topological invariant, the degree of the line bundle $L$.
Since $B$ is uniform, this implies 
\beq\label{B}
B=\frac{2\pi n}{e\area(\Sigma)}
\eeq
which, for the case $\Sigma=S^2$ with the round metric, coincides with the
celebrated Dirac quantization condition (one can interpret $B$ as
being the uniform field produced by a magnetic monopole placed at the
centre of the sphere). Now, given any
section $\psi\in\Gamma(L)$, we can define 
$\wh{\psi}=-i\psi\vol_\Sigma\in\Omega^2(L)$. Then
\bea
\ip{\dstar^{\nabla}\wh{\psi},\d^{\nabla}\psi}
=\ip{\wh\psi,\d^{\nabla}\d^{\nabla}\psi}
=\ip{-i\psi\vol_\Sigma,-ieB\psi\vol_\Sigma}
=eB\|\psi\|^2.
\eea
But, by Cauchy-Schwarz,
\bea
\ip{\dstar^{\nabla}\wh{\psi},\d^{\nabla}\psi}\leq
\|\dstar^{\nabla}\wh{\psi}\|\|\d^{\nabla}\psi\|
=\|\d^{\nabla}\psi\|^2
=\ip{\psi,\Delta^\nabla\psi}.
\eea
Hence,
\beq\label{bound}
\ip{\psi,\Delta^\nabla\psi}\geq eB\|\psi\|^2,
\eeq
whence $\lambda_0\geq eB$ as claimed.
Formula (\ref{e0}) is equivalent to the statement that the
topological lower energy bound (\ref{bound})
is attained, which, in turn, is equivalent to the statement that there
exists a nonzero section $\psi\in\Gamma(L)$
with
\beq\label{ladder}
*\d^{\nabla}\psi=i\d^{\nabla}\psi,
\eeq
because equality holds in the Cauchy-Schwarz inequality if and only
if $\dstar^{\nabla}\wh{\psi}=c\d^\nabla\psi$ for some $c>0$, and
$\|\dstar^{\nabla}\wh{\psi}\|=\|\d^\nabla\psi\|$, so $c=1$.
Perhaps a direct proof that (\ref{ladder}) has a nontrivial
solution is possible,
but we shall instead determine $\lambda_0$ by an indirect argument.

It is convenient henceforth to allow $\nabla$ to denote
a general metric connexion on $(L,h)$, and denote by $\nabla_0$ any
metric connexion
with uniform $B$ (for $\Sigma\neq S^2$, 
such connexions are not unique).
Consider the variational problem which assigns to a section $\phi$ of
$L$ and a connexion $\nabla$ the energy
\beq
E(\phi,\nabla)=\frac12\|\d^\nabla\phi\|^2+\frac12\|iF^\nabla\|^2+
\frac18\|\tau-h(\phi,\phi)\|^2
\eeq
where $\|\cdot\|$ denotes $L^2$ norm and $\tau>0$ is a positive constant.
This is the abelian Higgs model on $\Sigma$ and was studied (in a rather more
general setting) by Bradlow \cite{bra}. The field equations are obtained by
demanding that
\beq
\left.\frac{d\: }{dt}E(\phi_t,\nabla_t)\right|_{t=0}=0
\eeq
for all smooth variations of $\phi,\nabla$. Defining $\eta=\cd_t\phi_t|_{t=0}
\in\Gamma(L)$ and $\alpha=i\cd_t\nabla_t|_{t=0}\in\Omega^1(\Sigma)$, we see
that
\beq\label{first}
\left.\frac{d\: }{dt}E(\phi_t,\nabla_t)\right|_{t=0}
=\ip{\dstar^\nabla\d^\nabla\phi,\eta}-\ip{j_\phi,\alpha}
+\ip{i\dstar F^\nabla,\alpha}
-\frac12\ip{(\tau-h(\phi,\phi))\phi,\eta}
\eeq
where $\ip{\cdot,\cdot}$ denotes $L^2$ inner product,
$j_\phi$ is
the ``supercurrent'' one form
\beq
j_\phi(X)=h(\nabla_X\phi,i\phi),
\eeq
and $\dstar=-*\d*$ is the $L^2$ adjoint of $\d$.
 So $(\phi,\nabla)$ is
a critical point of $E$ if and only if
\beq
\dstar^\nabla\d^\nabla\phi=\frac12(\tau-h(\phi,\phi))\phi,
\qquad
\dstar(iF^\nabla)=j_\phi.
\eeq
Hence $(0,\nabla)$ is a critical point of $E$ for all
$\tau>0$ provided $\dstar F^\nabla=0$,
that is, provided $B$ is constant. So $(0,\nabla_0)$ is 
a critical point of $E$ for all $\tau>0$.

Let us consider how the stability properties of the 
critical point $(0,\nabla_0)$
depend on $\tau$. To determine whether a critical point of $E$ is
stable, we compute the second variation of $E$ about that critical point
\cite{atibot,horrairaw}.
So, let $(\phi_{s,t},\nabla_{s,t})$ be a smooth two-parameter variation
of $(0,\nabla_0)$, with infinitesimal variations $\eta=\cd_t\phi_{s,t}|_{(0,0)},
\nu=\cd_s\phi_{s,t}|_{(0,0)}\in\Gamma(L)$ and 
$\alpha=i\cd_t\nabla_{s,t}|_{(0,0)},\beta=i\cd_s\nabla_{s,t}|_{(0,0)}\in\Omega^1(\Sigma)$. Then, from (\ref{first}) we have
\beq
\left.\frac{\cd^2 E(\phi_{s,t},\nabla_{s,t})}{\cd s\cd t}\right|_{s=t=0}
=\ip{\dstar^{\nabla_0}\d^{\nabla_0}\nu,\eta}+\ip{\dstar\d\beta,\alpha}
-\frac\tau2\ip{\nu,\eta}.
\eeq
The critical point $(0,\nabla_0)$ is stable if the associated
quadratic form on $\Gamma(L\oplus T^*\Sigma)$,
\beq
Q(\eta,\alpha)=\ip{(\dstar^{\nabla_0}\d^{\nabla_0}-\frac\tau2)\eta,\eta}
+\ip{\dstar\d\alpha,\alpha}
\eeq
is non-negative.
Clearly $(0,\nabla_0)$ is stable against all variations of $\nabla$,
but is stability against variations of $\phi$ only while
$0<\tau\leq 2\lambda_0$, becoming unstable when $\tau>2\lambda_0$.

Now, it is known from work of Bradlow \cite{bra} that for all
\beq\label{tau0}
\tau>\tau_0=\frac{4\pi n}{\area(\Sigma)}
\eeq
the global minimum of $E$ is attained by a $n$-vortex solution (a
certain section $\phi$ and connexion $\nabla$ satisfying 
a first order system of PDEs, called
Bogomol'nyi equations, which imply the field equations). Furthermore,
in the limit that $\tau\ra\tau_0$ from above, these vortex solutions
converge to a uniform solution $(0,\nabla_0)$. 
Hence, $(0,\nabla_0)$ is
stable precisely at $\tau=\tau_0$ (since it globally minimizes $E$), but
becomes unstable for $\tau>\tau_0$ (since the lower energy $n$-vortex branch
bifurcates off at $\tau=\tau_0$). Comparing with our linear
stability analysis, we deduce that $\tau_0=2\lambda_0$. But recall
that the quantum Hamiltonian of interest is $H=\frac{1}{2m}\Delta_0$,
whose lowest eigenvalue is thus
\beq
E_0=\frac{1}{2m}\lambda_0=\frac{\tau_0}{4m}.
\eeq
Combining this with (\ref{tau0}) and (\ref{B}) gives the formula claimed 
(\ref{e0}). As an aside, we note that, 
since the bound (\ref{bound}) is attained by the 
ground state wavefunction, it must satisfy (\ref{ladder}).
 This reduces the problem of constructing
the ground state wavefunction to solving a first order linear PDE.
It would be interesting to see whether recent work by Manton and Romao
on the geometry of vortices in the limit $\tau\ra\tau_0$ yields any
useful information about this ground state \cite{manrom}.

\subsection*{Acknowledgements} This work was partially supported by the UK 
Engineering and Physical Sciences Research Council.
The author wishes to thank Derek Harland for useful correspondence.

\end{document}